# Antiadiabatic Small Polaron Formation in the Charge Transfer Insulator ErFeO$_3$

Ye-Jin Kim,[1] Jocelyn L. Mendes,[1] Young Jai Choi,[2] and Scott K. Cushing[1,*]

[1] Division of Chemistry and Chemical Engineering, California Institute of Technology, Pasadena, California 91125, United States

[2] Department of Physics, Yonsei University, Seoul, 03722, Republic of Korea

*Correspondence and requests for materials should be addressed to S.K.C. (email: scushing@caltech.edu)

**ABSTRACT**

Small polaron formation is dominant across a range of condensed matter systems. Small polarons are usually studied in terms of ground-state transport and thermal fluctuations, but small polarons can also be created impulsively by photoexcitation. The temporal response of the lattice and local electron correlations can then be separated, such as with transient XUV spectroscopy. To date, photoexcited small polaron formation has only been measured to be adiabatic. The reorganization energy of the polar lattice is large enough that the first electron-optical phonon scattering event creates a small polaron without significant carrier thermalization. Here, we use transient XUV spectroscopy to measure antiadiabatic polaron formation by frustrating the iron-centered octahedra in a rare-earth orthoferrite lattice. The small polaron is measured to take several picoseconds to form over multiple coherent charge hopping events between neighboring Fe$^{3+}$–Fe$^{2+}$ sites, a timescale that is more than an order of magnitude longer compared to previous materials. The measured interplay between optical phonons, electron correlations, and on-site lattice deformation give a clear picture of how antiadiabatic small polaron transport would occur in the material. The measurements also confirm the prediction of the Holstein and Hubbard-Holstein model that the electron hopping integral must be larger than the reorganization energy to achieve antiadiabaticity. Moreover, the measurements emphasize the importance of considering dynamical electron correlations, and not just changes in the lattice geometry, for controlling small polarons in transport or photoexcited applications.

**INTRODUCTION**

The interaction between charge carriers and a polar lattice can lead to charge localization by a local lattice deformation which is referred to as polaron formation. Polarons are of major interest for semiconducting, superconducting, and insulating materials because they dominate transport properties[1-4]. In the strong-coupling limit, sublattice-sized small (Holstein) polarons[5,6] reconfigure electronic states to control surface reactivity[7,8], ion mobility[9,10], and high-temperature superconductivity[11]. Understanding how strong electron-phonon coupling versus electron correlations control polaron formation is important from both a ground-state materials and photodynamics perspective.

Optical excitation can generate electron and hole polarons by photodoping the lattice[12–17]. To date, all photoexcited small polarons are measured to form within the first electron-optical phonon scattering event, even when the charge carrier has excess energy above the band edge[16,18–23]. Within the Hubbard-Holstein model for correlated electron materials, this indicates that the reorganization energy (electron-phonon coupling strength) of the small polaron is larger than the hopping integral between metal sites[24]. The hundreds of femtoseconds small polaron formation is termed adiabatic as it occurs without significant thermalization. In this case, the photoexcited small polaron formation is almost an impulsive excitation. Through fluctuation-dissipation theory, the resultant photodynamics are therefore directly related to transport properties, as recently demonstrated in measurements of $CuFeO_2$.

In this study, we measure that the distorted $FeO_6$ octahedron in erbium iron oxide ($ErFeO_3$), a rare-earth orthoferrite, leads to antiadiabatic small polaron formation using transient XUV spectroscopy. Photoexcitation of a ligand-to-metal charge transfer transition creates an axially elongated $FeO_6$ octahedra. Multiple coherent charge hopping events are then measured between neighboring $Fe^{3+}$–$Fe^{2+}$ sites. The coherent charge hopping is associated with an optical phonon mode that modulates the distance between the neighboring Fe–O–Fe bonds. The charge hopping continues until the photoexcited carriers thermalize and small polaron formation occurs on picoseconds timescale, as is characteristic of an antiadiabatic interaction. The polaron formation time is an order of magnitude longer than previously measured[16,18,20-23]. The lattice reorganization energy estimated from the experiments is also smaller than the magnitude of the hopping integral, roughly estimated by the Fe $d$-orbital conduction band width[25], confirming previous Holstein and Hubbard-Holstein model predictions for the conditions of antiadiabatic polaron formation.

Overall, our measurements confirm that dynamical electron correlations play a significant role in controlling small polaron formation in correlated electron materials and must be considered beyond just the lattice reorganization energy[26]. This result is in contrast to a variety of previous studies in other Fe–O materials where changes to defects, dopants, spin, and symmetry always lead to a similar, adiabatic small polaron formation time, suggesting that knowledge of the reorganization energy alone was sufficient[16,18–23]. The measured interplay between optical phonons, electron correlations, and local lattice deformation provide a clear picture into how antiadiabatic ground-state polaron transport would occur. The findings are therefore important for large classes of small polaron limited, or small polaron enabled, applications while also giving new fundamental insight into how strong electron-phonon coupling manifests in correlated electron systems.

## RESULTS AND DISCUSSION

$ErFeO_3$ is an antiferromagnetic, charge transfer insulator that crystallizes in an orthorhombically distorted perovskite structure with space group *Pbnm*, as confirmed by X-ray diffraction (**Supplementary Note 1**, **Figure S1**, and **Table S1**)[27] and depicted in **Figure 1a**. It belongs to the rare-earth orthoferrite family $RFeO_3$ (where R is a trivalent rare-earth cation) and features corner-shared $FeO_6$ octahedra that are orthorhombically-distorted and -rotated about the [001] axis[28]. $ErFeO_3$ exhibits rich magnetic properties arising from the magnetic transitions between the $3d$ and $4f$ electrons of $Fe^{3+}$ and $Er^{3+}$ ions, respectively. These properties include a high Néel temperature[29], large magnetoelectric coupling[30], and ultrafast optical control of spins[31,32], making it a potential room-temperature multiferroic candidate. Here, we use this control of the frustrated lattice versus electron correlations to explore the structural factors that determine small polaron formation.

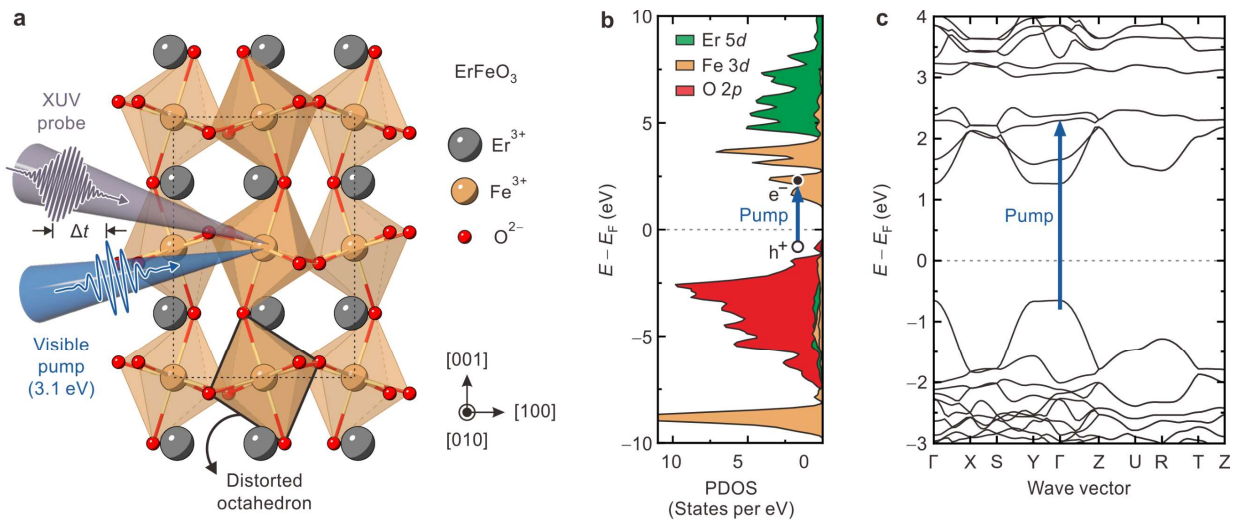

**Figure 1. Electronic and lattice structure of $ErFeO_3$. a,** Crystal structure of $ErFeO_3$. The crystal has a distorted perovskite structure with orthorhombic symmetry. Corner-sharing $FeO_6$ octahedra are distorted due to the presence of an $Er^{3+}$ ion. $ErFeO_3$ is photoexcited with a 50 fs, 3.1 eV pump pulse while XUV pulses probe photoinduced dynamics of the material. The unit cell of the crystal is indicated with a dotted box. **b,** Calculated total projected density of states (PDOS) of $ErFeO_3$. Photoexcitation generates holes (a white circle) and electrons (a black filled circle) in the O 2*p* and Fe 3*d* bands, respectively. $E_F$ is the fermi energy. **c,** Calculated band structure of $ErFeO_3$ along the high-symmetry *k*-points.

**Figures 1b** and **c** present the calculated electronic structures of the $ErFeO_3$ by using density-functional theory (DFT) with the Hubbard U correction (DFT+U). The calculations give a charge-transfer band gap of approximately 1.8 eV (calculation details in **Supplementary Note 2.1**, **Figure S2**), which is consistent with the measured charge-transfer absorption feature at $1.8 \pm 0.1$ eV (**Figure S3**). Within the approximations of DFT+U[33], the calculated projected density of states (PDOS) in **Figure 1b** has a valence band that is predominantly O $2p$ orbitals and a conduction band predominantly of lower-lying Fe $3d$ and higher-lying Er $5d$ orbitals[34,35]. The optical excitation used in these experiments, 3.1 eV, promotes electrons from the O $2p$ orbital to the unoccupied Fe $3d$ orbital through a ligand-to-metal charge transfer transition. The calculated band structure in **Figure 1c** along high-symmetry $k$-points reveals that the valence band maximum and the conduction band minimum are flat from Γ- to Y-points of the Brillouin zone, indicative of strong electron correlations[36], similar to other rare-earth orthoferrites[37,38].

Transient XUV spectroscopy is used to measure electron dynamics following a ligand-to-metal charge transfer photoexcitation. Time-resolved differential absorption spectra are collected at the Fe $M_{2,3}$ edges around 54 eV which corresponds to the $3p_{3/2,\ 5/2}$ core states to $3d$ valence state transition. Photoexcitation is with a 50 fs, 3.1 eV (400 nm) pulse at a pump fluence of 3 mJ/cm$^2$ under ambient conditions. The differential absorption signal, $\Delta A$, is calculated as $\Delta A = \log_{10}(R_{off}/R_{on})$, where $R_{off}$ and $R_{on}$ are the XUV reflectance spectra with pump beam blocked and unblocked, respectively. The high harmonic generation technique creates the XUV probe pulse (upper panel in **Figure 2a**) from a few-cycle white light pulse in an argon gas medium (experimental details in **Supplementary Note 3** and **Figure S4**)[39]. By employing a grazing incidence angle of 10°, the XUV has a penetration depth of approximately 2 nm, providing an element-specific and surface-sensitive geometry[39,40]. In the measurement of small polarons in $Fe_2O_3$ in the past, bulk versus surface measurements only showed variations in the 30% range[18], smaller than the order of magnitude change in small polaron formation time measured here.

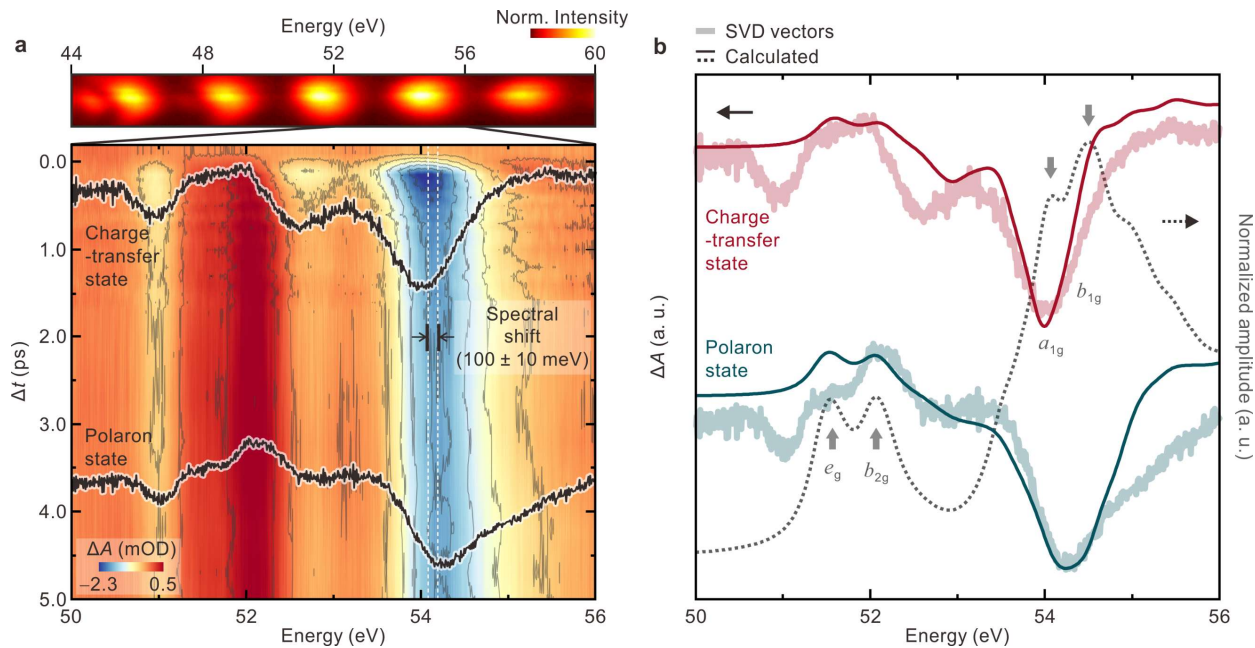

**Figure 2. Photoinduced structural distortion and small polaron formation. a,** (Upper) high-order harmonic profile for the XUV probe pulses. Color scale corresponds to the normalized harmonic intensity. (Lower) two-dimensional false color plot of time-resolved differential absorption spectra obtained at pump fluence of 3 mJ/cm². The color scale corresponds to differential absorbance ($\Delta A$). Vectors for singular value decomposition (SVD) are overlaid, representing (upper) the charge-transfer and (lower) the small polaron states. The spectral blue-shift of the main absorption feature at ~54 eV, typically indicative of small polaron formation, is measured as 100 ± 10 meV. **b,** Calculated differential absorption spectra compared to the primary SVD vectors. For the charge-transfer state, the FeO₆ octahedron is axially elongated due to an optical phonon within our temporal resolution, giving rise to crystal field splitting, as shown in the calculated excited-state absorption spectra of $Fe^{2+}$ (dotted line). State-filling effects were incorporated for the photoexcited electrons and holes. For the polaron state, six Fe–O bonds in the octahedron are expanded with the overall enlargement of the unit cell.

The time-resolved differential absorption spectra are shown in **Figure 2a** as a pseudo-color plot (see **Figure S5** for the temporal lineout spectra). The first 5 ps after photoexcitation are plotted, after which the dynamics become steady state out to the 1 ns measurement limit of our instrument (**Figure S6**). The X-ray transition Hamiltonian for Fe edges is dominated by angular momentum components[41]. The increase and decrease in absorption in the XUV spectrum therefore do not directly relate to electron and hole energies discussed in previous XUV studies, but instead relates to the change in overall Fe site occupation (oxidation state), the presence of phonon modes, and other structural excitations such as small polarons[41,42].

The spectral evolution exhibits two distinctive vectors and timescales within the 5 ps time frame, which can be confirmed by singular value decomposition[20,43] (see overlays in **Figure 2a**).

An excited-state approximation to the Bethe-Salpeter equation is used to interpret the spectral contributions[44,45]. Briefly, the OCEAN (Obtaining Core Excitations from the *Ab initio* electronic structure and the NIST BSE solver) code is modified using an adiabatic approximation of the photoexcited electronic states or lattice distortions to calculate the change in the core-valence excitons that make up the XUV spectra (calculation details in **Supplementary Note 2.2**)[39,41,42]. The initial photoexcited charge transfer state creates four distinct bands positioned at 51.6, 52.1, 54.1, and 54.5 eV when compared to the $Fe^{3+}$ ground state, **Figure 2b**, which can be assigned to $e_g$, $b_{2g}$, $a_{1g}$, and $b_{1g}$ multiplets, respectively[46]. A local elongation of the Fe–O bond in the axial direction, and a contraction of the equatorial in-plane bonds from the Jahn-Teller distortion (**Figure S7**), must also be included. The interatomic distance between neighboring Fe sites decreases because of this distortion[47]. Next, at a pump-probe time delay after $\Delta t = 1$ ps, the main spectral feature at 54 eV blue-shifts. This spectral shift by $100 \pm 10$ meV, which also results in the change in the $e_g$ and $b_{2g}$ spectral amplitudes over time, corresponds to the formation of the small polaron[16,20,23]. This is modeled as an overall expansion of the Fe–O bonds, and good agreement exists between theory and experiment in both cases.

The kinetics of the small polaron formation are determined using the assigned spectral features. A multivariate regression analysis using the two spectra is depicted in **Figure 3a**. A three-state sequential kinetic model is used for the best fit as outlined in **Figure 3b** (fitting details in **Supplementary Note 4**), including the 50-fs photoexcitation pulse. The kinetic model is consistent with past small polaron formation studies[20,21] so that relative timescales can be compared. The initial charge transfer photoexcitation thermalizes on the order of $\tau_1 = 250 \pm 80$ fs, but the small polaron formation is delayed by $\tau_2 = 2.3 \pm 0.3$ ps with the incorporation of an intermediate step. This is characteristic of a balance between carrier thermalization and small polaron formation through the coherent charge hopping in the antiadiabatic regime as discussed in the next paragraph.

Further insight into the small polaron formation process is given by the coherent oscillations in the transient XUV spectra. **Figure 3c** shows coherent oscillations for lineouts at 50.6 eV and 52.2 eV that represent $Fe^{3+}$ and $Fe^{2+}$, respectively, based on the spectral assignment (**Figure S7**) and phase coherences within each state, see **Figures S8** and **S9** for the full spectrum and analysis. The coherent oscillations have a periodicity of approximately 170 fs, corresponding to a frequency of 5.9 THz (196 cm$^{-1}$). The measured oscillation frequency can be correlated with a Raman-active optical phonon mode with $A_g$ symmetry that corresponds to a rotation of the $FeO_6$ octahedra between the $Er^{3+}$ ions[28,48], see the schematic in **Figure 3d**. The optical phonon mode leads to coherent charge hopping between neighboring Fe sites by modulating the Fe–O–Fe bond length, as evidenced by the 180 degrees out-of-phase signal between the $Fe^{3+}$ and $Fe^{2+}$ signals (**Figure 3d**). The coherent charge hopping events in **Figure 3c** do not onset until significant carrier thermalization has occurred, matching the first timescale of the bi-exponential from **Figure 3a**. Similarly, small polaron formation does not complete in **Figure 3a** until after the multiple coherent charge hopping events decay in **Figure 3c**, matching the second timescale of the bi-exponential, designating the process as antiadiabatic.

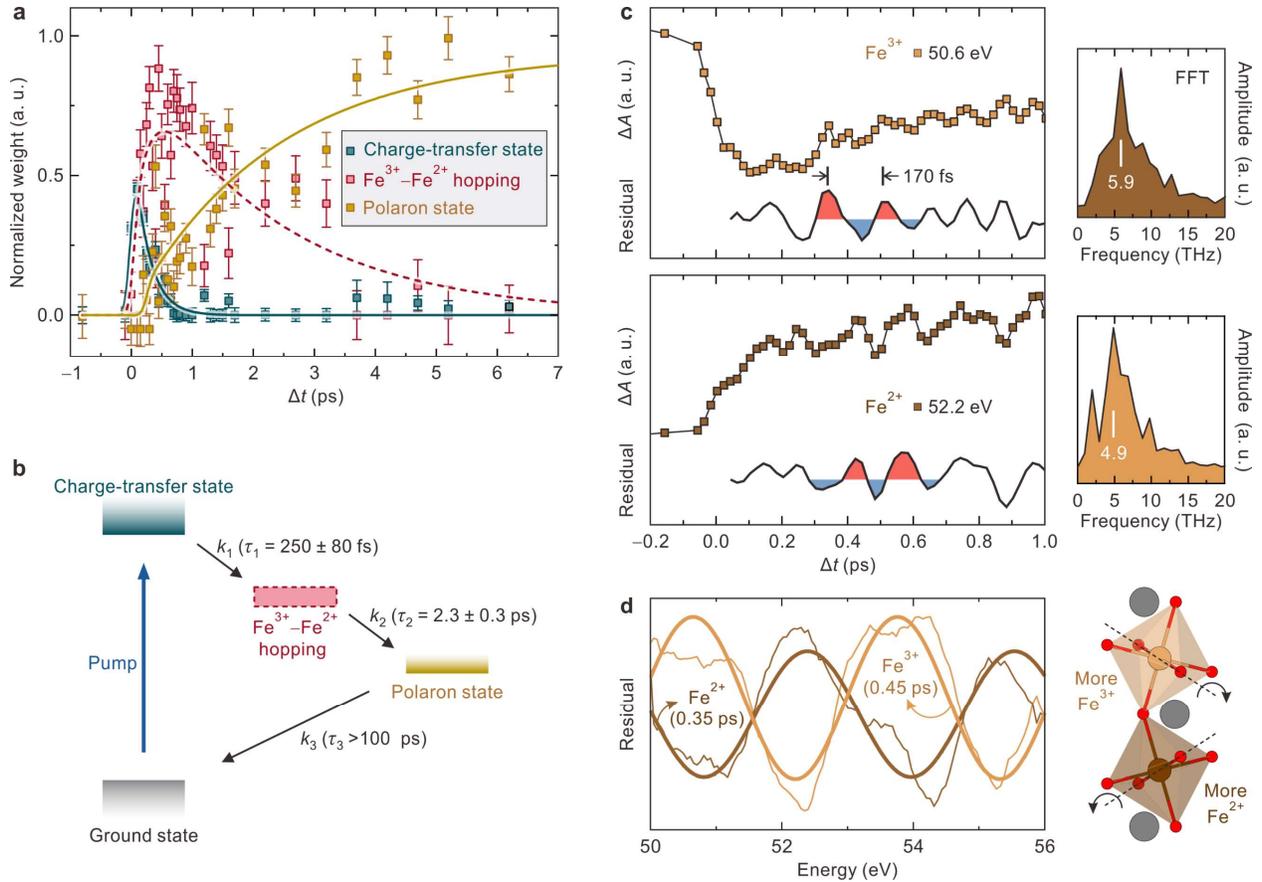

**Figure 3. Kinetic model of the relaxation dynamics and coherent hopping motions. a**, The time-dependent amplitude of the charge-transfer and small polaron states using a multivariate regression of the experimental data. Three-state model was employed to incorporate the intermediate step of the charge hopping events in antiadiabatic regime. The amplitudes are fitted with a rate equation model for small polaron formation (solid and dotted lines). **b,** Schematics of the three-state sequential kinetic model with the three time constants of $\tau_1$, $\tau_2$, and $\tau_3$. **c**. The coherent modulation of transient signals extracted at 50.6 and 52.2 eV, representing $Fe^{3+}$ and $Fe^{2+}$, respectively.. Oscillations from the initial electron–optical phonon scattering lead to coherent hopping between the photoexcited $Fe^{3+}$ and its $Fe^{2+}$ nearest neighbors. The fast Fourier transforms (FFT) are shown on the right. **d.** The out-of-phase oscillations between the $Fe^{3+}$ and $Fe^{2+}$ spectral features at $\Delta t$ between 0.3 to 0.6 ps indicative of the charge-hopping process that delays small polaron formation in the antiadiabatic process.

Transient XUV measurements therefore provide a detailed picture of antiadiabatic small polaron formation (**Figure 4**). The photoexcited electrons of the axially elongated $FeO_6$ octahedra thermalize through scattering with optical phonon modes, some of which correspond to $FeO_6$ octahedral rotations within the $Er^{3+}$ sub-lattice. The octahedral rotations modulate Fe–O–Fe bond lengths, leading to a periodic hopping between sites. Only after multiple electron-optical phonon scattering events does the electron hopping decay, corresponding to the creation of the small polaron. The small polaron does not lead to recombination within the hundreds of picoseconds timescale of the transient XUV measurement. The measured small polaron formation process is in contrast with the previously measured adiabatic mechanism, wherein the small polaron forms immediately within the first electron-optical phonon scattering events without significant carrier thermalization[16,18,20–23].

The Hubbard-Holstein model predicts that the adiabaticity of small polaron formation depends on the balance of the hopping integral between Fe sites and the small polaron reorganization energy. The reorganization energy is usually parameterized by the phonon frequency and deformation potential, balanced with the on-site interaction of the Fe. Previous studies on $\alpha$-$Fe_2O_3$[16,18–23] demonstrated that the blue-shift of the main XUV spectral feature can be related to the small polaron reorganization energy. In these studies, a spectral shift of ~1 eV corresponded to a formation energy of 0.4–0.6 eV, consistent with the theoretical predictions. Here, the measured spectral shift is $100 \pm 10$ meV (**Figure 2a** and **Figure S5**), which is one-tenth of the reported value for $\alpha$-$Fe_2O_3$. Given the difference between the adiabatic and antiadiabatic formulas for formation energies, the shifts can only be roughly compared, bounding the small polaron formation energy of $ErFeO_3$ as <100 meV, or <50 meV if the same XUV relation holds. In support of this, the polaron formation rate is also roughly one-tenth of that reported for $\alpha$-$Fe_2O_3$. The hopping integral ($t$) can be estimated as <200 meV by the bandwidth of the Fe 3$d$ orbital (~1 eV, **Figure 1b**) through the relationship $2dt$, where $d$ is the dimensionality of the system. Even when using upper bounds, the measurements confirm the Hubbard-Holstein prediction that the hopping integral must be greater than the reorganization energy for antiadiabatic small polarons to exist, and the coherent electron hopping suggests the need for a dynamic or renormalized electron-electron interaction parameter[26].

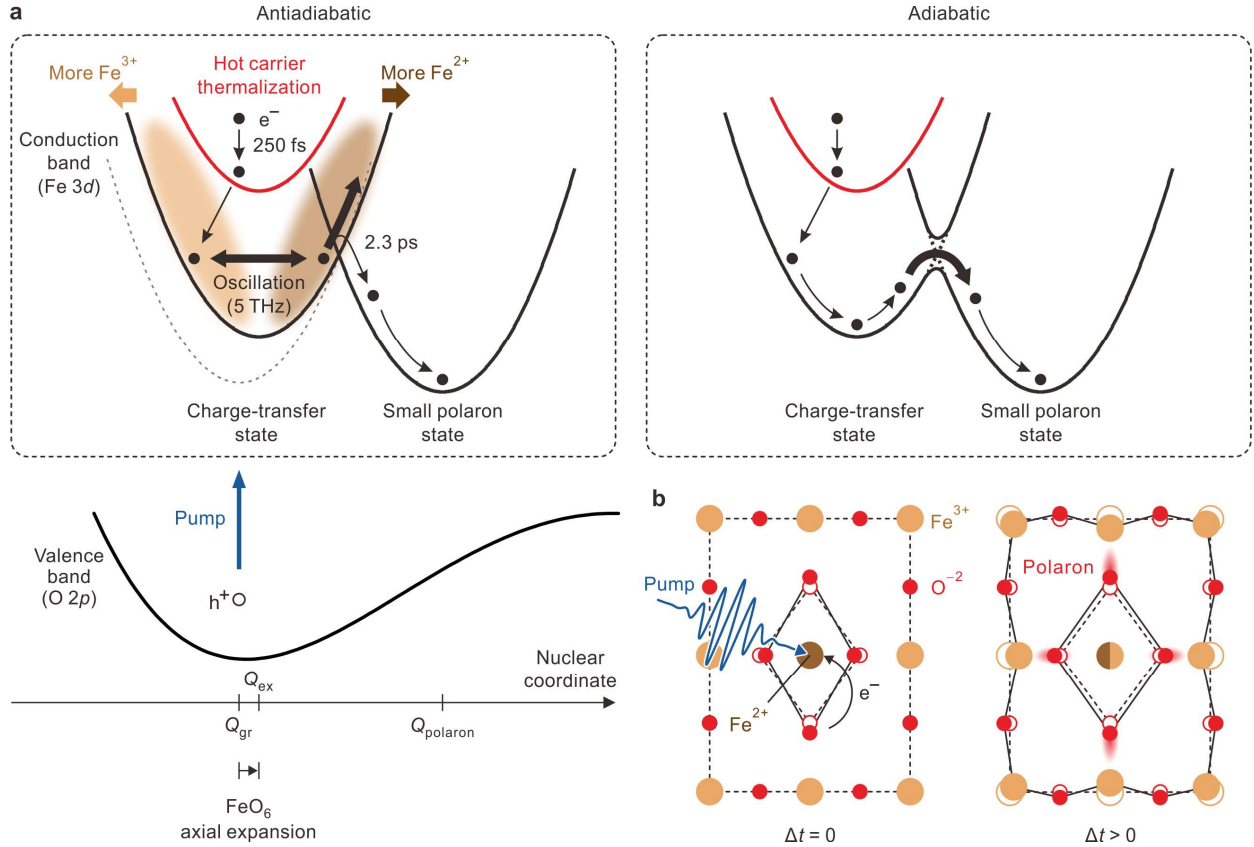

**Figure 4. Summary of the antiadiabatic photoinduced polaron formation dynamics in ErFeO$_3$. a,** Potential energy curves comparing small polaron formation in the antiadiabatic and adiabatic regimes. Upon photoexcitation, electrons instantaneously transfer from the O$^{2-}$ to Fe$^{3+}$ ions, generating photoexcited electrons and holes in the conduction and valence bands, respectively. In the charge-transfer state, the axial Fe–O bond is elongated, resulting in the splitting of the Fe 3$d$ band at a slightly shifted nuclear coordinate from the ground- to excited-state configurations ($Q_{gr} \rightarrow Q_{ex}$). Hot carriers in the high energy conduction band then thermalizes in hundreds of femtoseconds. In the antiadiabatic polaron formation of this work, coherent charge hopping between the Fe$^{3+}$ and Fe$^{2+}$ lasts a few picoseconds delaying small polaron formation at $Q_{polaron}$, while electron tunneling is dominant in the adiabatic pictures. **b,** Schematics of lattice deformation in ErFeO$_3$ at the charge-transfer ($\Delta t = 0$) and small polaron states ($\Delta t > 0$), where FeO$_6$ octahedron axially elongates upon photoexcitation and then expands equally in all axes during the small polaron formation.

Our measurements indicate that a complete picture of small polaron formation in correlated electron systems must include the balance of the charge hopping integral, here assisted by optical phonons, versus the lattice-based reorganization energy. Usually, a Holstein-like model is considered sufficient, focusing mainly on the reorganization energy, but we confirm that a full Hubbard-Holstein-like model needs to be used when the polaron formation energy approaches the electron bandwidth in correlated electron materials. The coherent charge hopping between Fe–Fe sites also confirms how antiadiabatic small polaron hopping would happen in the transport, requiring a coordination of optical phonons, electron correlations, and local lattice deformations at adjacent sites to allow site-to-site hopping. This is opposite to the adiabatic picture where polaron formation can occur through charge hopping *via* tunneling without the need for additional generated excess phonon modes[16,18,20–23].

## DATA AVAILABILITY

The data that support the findings of this study are available within the text including the Methods, and Supplementary information. Raw datasets related to the current work are available from the corresponding author on reasonable request.

## AUTHOR INFORMATION

**Corresponding Author**

**Scott K. Cushing** – Division of Chemistry and Chemical Engineering, California Institute of Technology, Pasadena, California 91125, United States

**Authors**

**Ye-Jin Kim** – Division of Chemistry and Chemical Engineering, California Institute of Technology, Pasadena, California 91125, United States

**Jocelyn L. Mendes** – Division of Chemistry and Chemical Engineering, California Institute of Technology, Pasadena, California 91125, United States

**Young Jai Choi** – Department of Physics, Yonsei University, Seoul, 03722, Republic of Korea

## ACKNOWLEDGEMENTS

Y.-J.K. is supported by the Liquid Sunlight Alliance (the U.S. Department of Energy, Office of Science, Office of Basic Energy Sciences, Fuels from Sunlight Hub under Award Number DE-SC0021266). J.L.M. acknowledges support by the National Science Foundation Graduate Research Fellowship Program under Grant No. 1745301. Any opinions, findings, and conclusions or recommendations expressed in this material are those of the author(s) and do not necessarily reflect the views of the National Science Foundation. The computations presented here were conducted in the Resnick High Performance Computing Center, a facility supported by the Resnick Sustainability Institute at the California Institute of Technology. We thank Won-Woo Park at Ulsan National Institute of Science and Technology for measuring the diffuse reflectance spectrum. We also thank Wonseok Lee at California Institute of Technology for providing a fruitful comments on the *ab initio* calculation.

**AUTHOR CONTRIBUTIONS**

Y.-J.K. performed the measurement with the initial support of J.L.M. Y.-J.K. analyzed the data and carried out the *Ab initio* calculations. Y.J.C. fabricated and characterized the specimen. S.K.C. supervised the project and provided guidance. Y.-J.K. and S.K.C. wrote the manuscript with input from all authors.

**COMPETING INTERESTS**

The authors declare no competing interests.